# Radiometry for Nighttime Sub-Cloud Imaging of Venus' Surface in the Near-InfraRed Spectrum

Brian M. Sutin, Anthony B. Davis, Kevin H. Baines, James A. Cutts, and Leonard I. Dorsky

Jet Propulsion Laboratory / California Institute of Technology,
4800 Oak Grove Drive, Pasadena, CA 91109, USA

**Abstract**

Does radiometry (e.g., signal-to-noise ratio) limit the performance of near-IR sub-cloud imaging of our sister planet's surface at night? It does not. We compute sub-cloud radiometry using above-cloud observations, an assumed ground temperature, sub-cloud absorption and emission modeling, and Rayleigh scattering simulations. We thus confirm both archival and recent studies that deployment of a modest sub-cloud camera does enable high-resolution surface imaging.

## 1. Introduction

### 1.1 Science context

Multiple flyby and orbital sensors have provided compelling evidence that Venus' surface emission can be imaged over the night side of our sister planet in the near-IR spectrum thanks to a few narrow windows in the $CO_2$ absorption spectrum. However, the effective resolution of these above-cloud observations is only on the order of 100 km due to the powerful blurring effect of the optically thick cloud layer between 50 to 70 km altitude.

These data have nonetheless opened far-reaching questions about Venusian geology, in particular about the possibility of recent or even currently active volcanism [0]. From there, it is a tantalizing idea to deploy a near-IR imaging sensor just below the cloud layer that could reveal surface geomorphology at the ~10 m scale. Such data would undoubtedly revolutionize Venusian geology [P.K. Byrne, pers. com.].

### 1.2 Challenge at hand

Venus is shrouded in clouds above 50-km altitude. Finer resolution imaging than the ~100 km achievable from orbit requires an optical instrument below the cloud cover. What would such an instrument see? Three processes degrade the imaging of surface features: atmospheric scattering, camera performance, and turbulence (in that it affects seeing & platform stability).

We conclude that the first two of these effects do not preclude observing the ground from 47-km altitude at resolutions of 20-m FWHM (10-m sampling) in bands at or above 1 micron or 150-m FWHM (75-m sampling) below 1 micron.

### 1.3 Outline

In the following Section, we collate a list of processes that matter for the radiometry of a camera on a just-below-cloud platform. In Section 3, we break down the radiometric estimation into 6 steps. Section 4 covers the aspects of instrument performance that matter for surface imaging with a sub-cloud camera. Finally, in Section 5, we state the conclusions of our study, putting them in the context of previous investigations.

## 2. Physical Processes & Radiometry

The structure of the Venus atmosphere is illustrated in *Figure 1*.

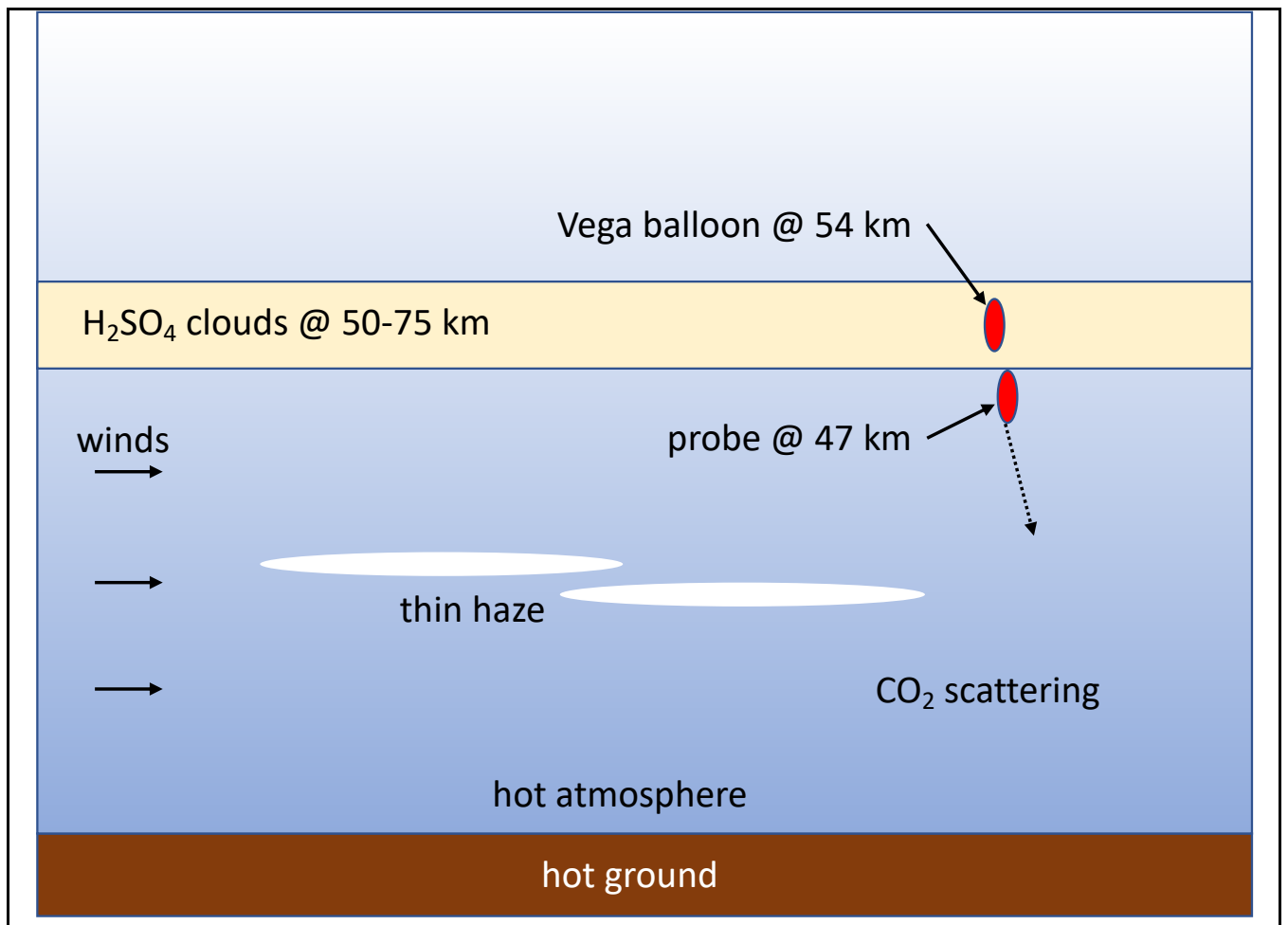

*Figure 1. Illustration of the components of the Venus atmosphere*

At nighttime, the science signal of interest is the ground emitting thermal radiation as a ≈735 K blackbody. This signal is then contaminated by hot atmosphere emission and in-scattering and reduced by extinction (absorption plus scattering) as it travels upwards through the atmosphere. To date, the best observations of the nighttime ground emission have been done from orbit. The fate of a science signal photon to orbit is

- Emission from the surface:
  - Intensity is modulated by emissivity and ground temperature (science).
  - Temperature (and thus intensity) is modulated by the ground elevation (lower is hotter). However, elevation is known from Magellan radar measurements, so we assume that this effect is compensated for in post-processing.
- Emission from hot atmosphere is added to the ground signal:
  - This has been modeled from 0.97 to 1.22 µm [1].
  - Since absorption of cold $CO_2$ and emission of hot $CO_2$ are two sides of the same process, the Meadows and Crisp model [1] can be heuristically extended down to 0.8 µm.
  - **This atmospheric emission adds shot noise to the science signal.**
- Rayleigh scattering from $CO_2$ creates a near-uniform background but does not degrade the un-scattered signal point spread function (PSF) [2]:
  - **This haze adds more shot noise to the science signal.**

- Absorption of cold $CO_2$ in the atmosphere creates the wavelength dependent bands or windows:
  - The widths of the absorption bands vary with altitude. This effect is currently ignored, with only the worst case at 47-km altitude being considered. Since little $CO_2$ is above 47 km, the window widths are the same as from orbit.
  - **Absorption reduces signal**
- The low-altitude $H_2O$, 30-50 km haze layers, and the 50-70 km $H_2SO_4$ clouds **absorb and scatter the signal:**
  - The resolution in orbit is reduced to c. 100 km [3-5].
  - The low-level $H_2O$ absorbs a small amount over 1.05-1.2 µm, but is highly variable.
  - The $H_2SO_4$ clouds transmit 10-20% of the light impinging on their base [3,4].
  - Effect of the 30-km haze layers is analyzed in [6,7].
  - Separating the haze and cloud layers is difficult, but also not required.

Given the above, we want to know the science signal and SNR for an instrument below the cloud deck, which we nominally choose as 47 km. While 47 km is the worst below-cloud case for SNR and ground resolution, lower altitudes have increasing thermal and pressure challenges to be overcome by the hardware.

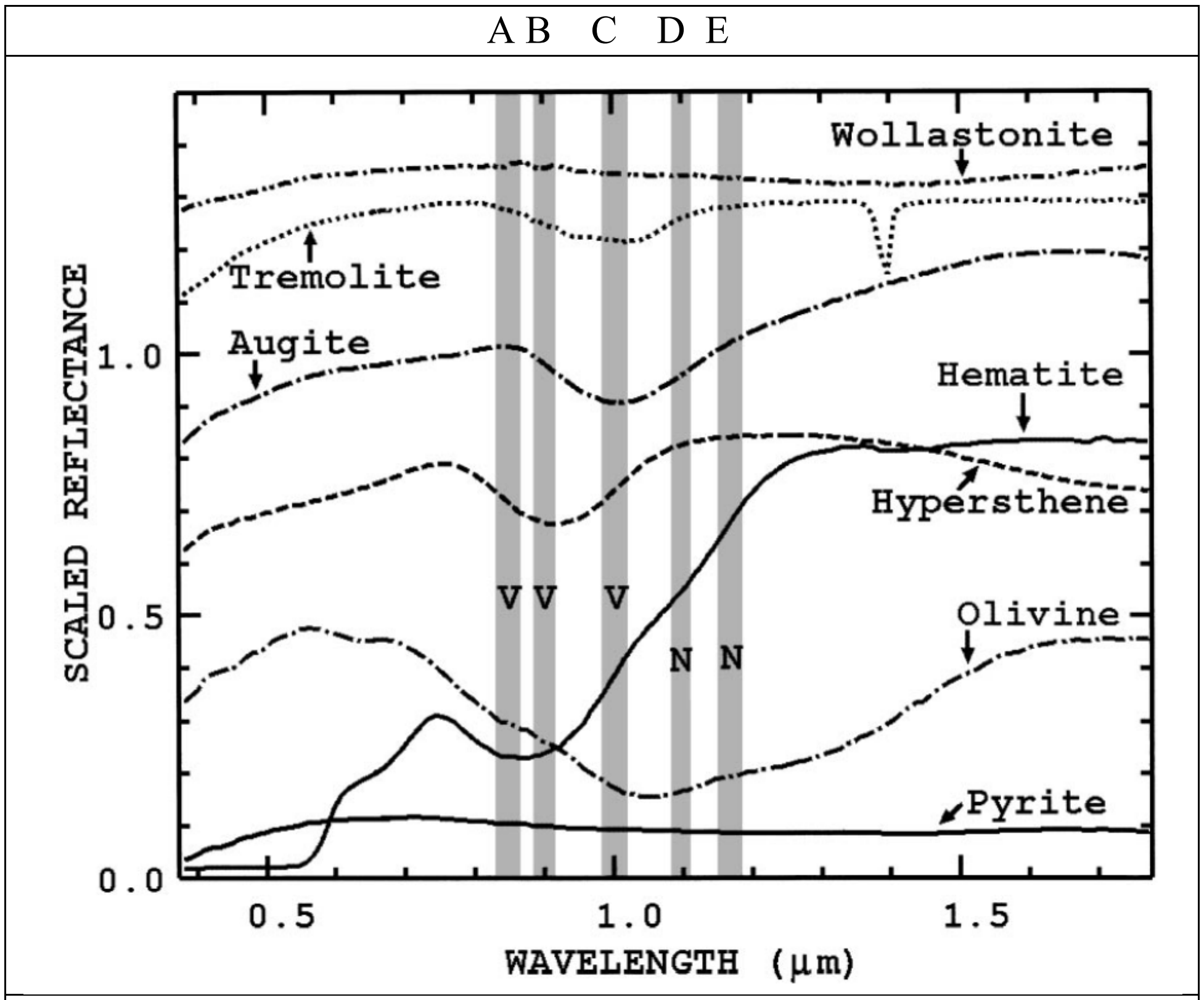

*Figure 2. Adapted from [4]. The reflectance shown here is (1–ε). Grey bars are the atmospheric absorption windows where the surface can be observed from orbit. The ABCDE labels at the top are designations of the atmospheric windows for this white paper only.*

# 3. Radiometry

## 3.1 Step 1 – Above-cloud data

The entire region of interest, ~0.8-1.2 µm, has not to our knowledge been observed as a single data set. We assemble a single data set by merging three data sets. For 0.8-1.02 µm, we use the Cassini/VIMS data from Baines et al. [8], hereafter referred to as 'Baines'. This data set extends to 1.05 µm, but the data after ~1.02 µm is highly affected by the instrument upper bandpass cutoff. For 1.06-1.22 µm, we use VEX/SPICAV-IR data from Bèzard et al. [9], hereafter referred to as 'Bèzard'. For the missing segment from ~1.02-1.06 µm, we use 0.97-1.33-µm AAT/IRIS data from Meadows & Crisp [1], hereafter referred to as 'Meadows'. The above-cloud data is digitized at a sampling of 2 nm. The digitized "fits" are shown in *Figure 3.*

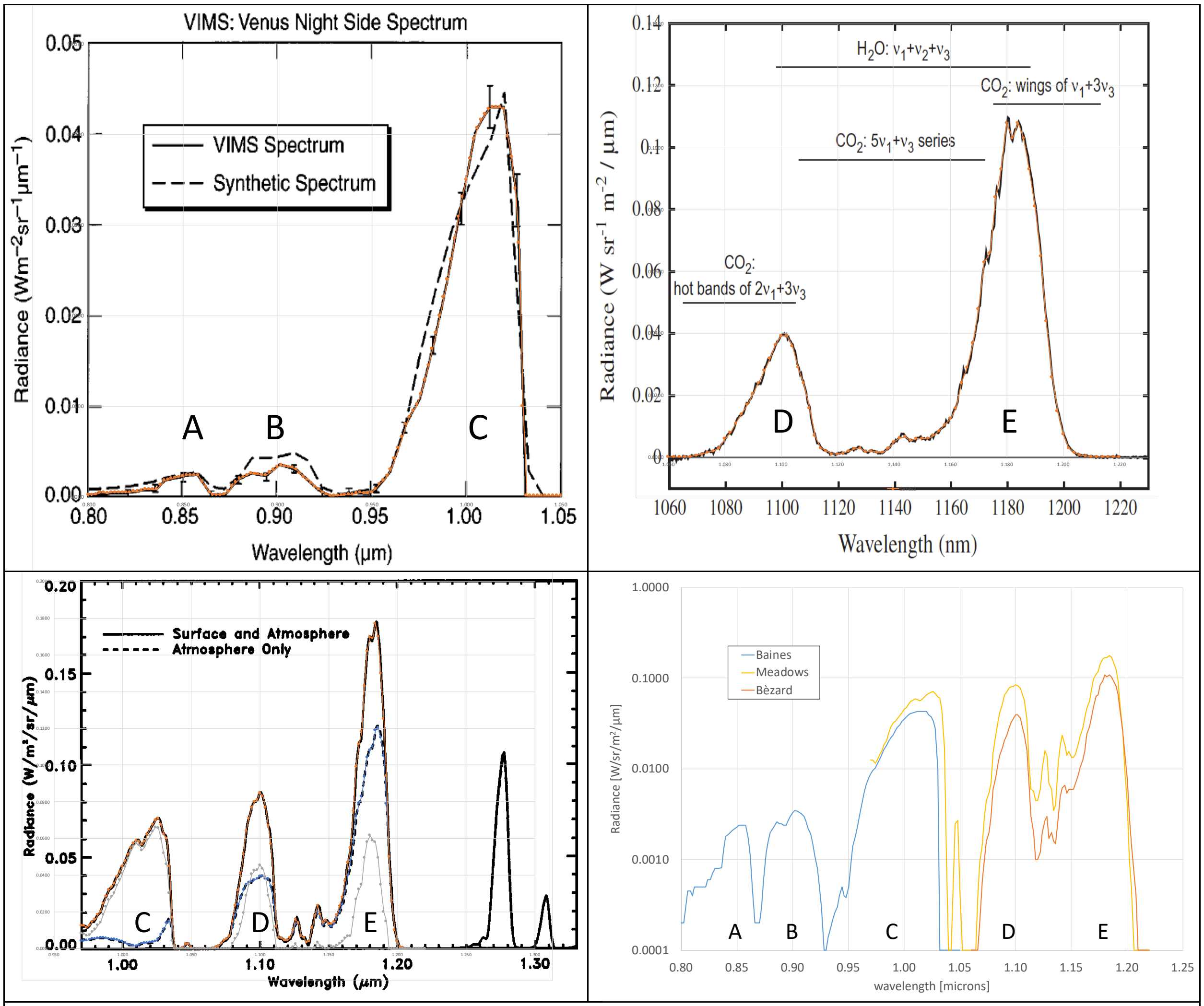

*Figure 3. Upper left [4] (Baines); upper right [5] (Bèzard); lower left [1] (Meadows; grey overlay is surface only); lower right is all source data on a single plot. The "ABCDE" labels defined in Figure 2 are specific to this work and are only to guide the eye on plots to the five atmospheric windows. Orange with markers is the digitization overlay.*

**3.2 Step 2 – Consistent radiometry**

The three data sets have mismatched radiometry. This could be from any number of reasons, including calibration offsets, varying cloud opacity, or differing ground terrain altitude. The actual absolute calibration is not important, as it will ultimately get absorbed into our estimate of cloud deck losses.

First, we accept the calibration of Bèzard from 1.06-1.22 µm. We then find a scaling (56.2%) of Meadows that minimizes the mean-square differences from Bèzard over 1.06-1.22 µm. Next, we find a scaling (75.2%) of Baines that minimizes the mean-square differences from Meadows over 0.97-1.02 µm. The final data set is chosen with the transition wavelengths where the radiances of the scaled data sets are equal. This results in Baines from 0.8-1.016 µm, Meadows from 1.016-1.06 µm, and Bèzard from 1.06-1.22 µm. The resulting combined above-cloud radiometry is shown in the upper left panel of *Figure 4.*

**3.3 Step 3 – Atmospheric emission**

The above-cloud data emission source is a combination of the hot surface and the hot atmosphere immediately above the surface. For the purposes of ground imaging, the hot atmosphere is a noise source to be removed. Meadows provides the atmospheric emission radiance from 0.97-1.3 µm, leaving 0.8-0.97 µm unknown. However, as emission and absorption are highly correlated, we can use the known absorption to estimate the emission.

The cores of bands A and C are dominated by emission near the ground; a significant span of wavelengths in the cores of these bands can be well-fit by a blackbody curve and some constant emissivity. We approximate the hot surface, hot atmosphere, and cloud losses as a reduced blackbody, 706.4 K, 16.5% "emissivity" (*Figure 4*, upper right panel). Keep in mind that this blackbody curve is only a functional fit, and the fit temperature does not correspond to any physical temperature.

Dividing the above-cloud data by the blackbody gives the absorption (*Figure 4*, lower left panel). The atmospheric emission from Meadows is highly correlated, as expected (*Figure 4,* lower left panel). The only significant deviation is in the core of band D, which we hereafter ignore as potentially interesting but ultimately off-topic. The final estimated atmospheric emission is shown in *Figure 4*, lower right panel. The estimate uses the absorption-based estimate below 0.996 µm as well as at points where the Meadows data was too small to be discernable on the published plot. We then separate the ground emission from hot atmosphere (*Figure 5,* upper left).

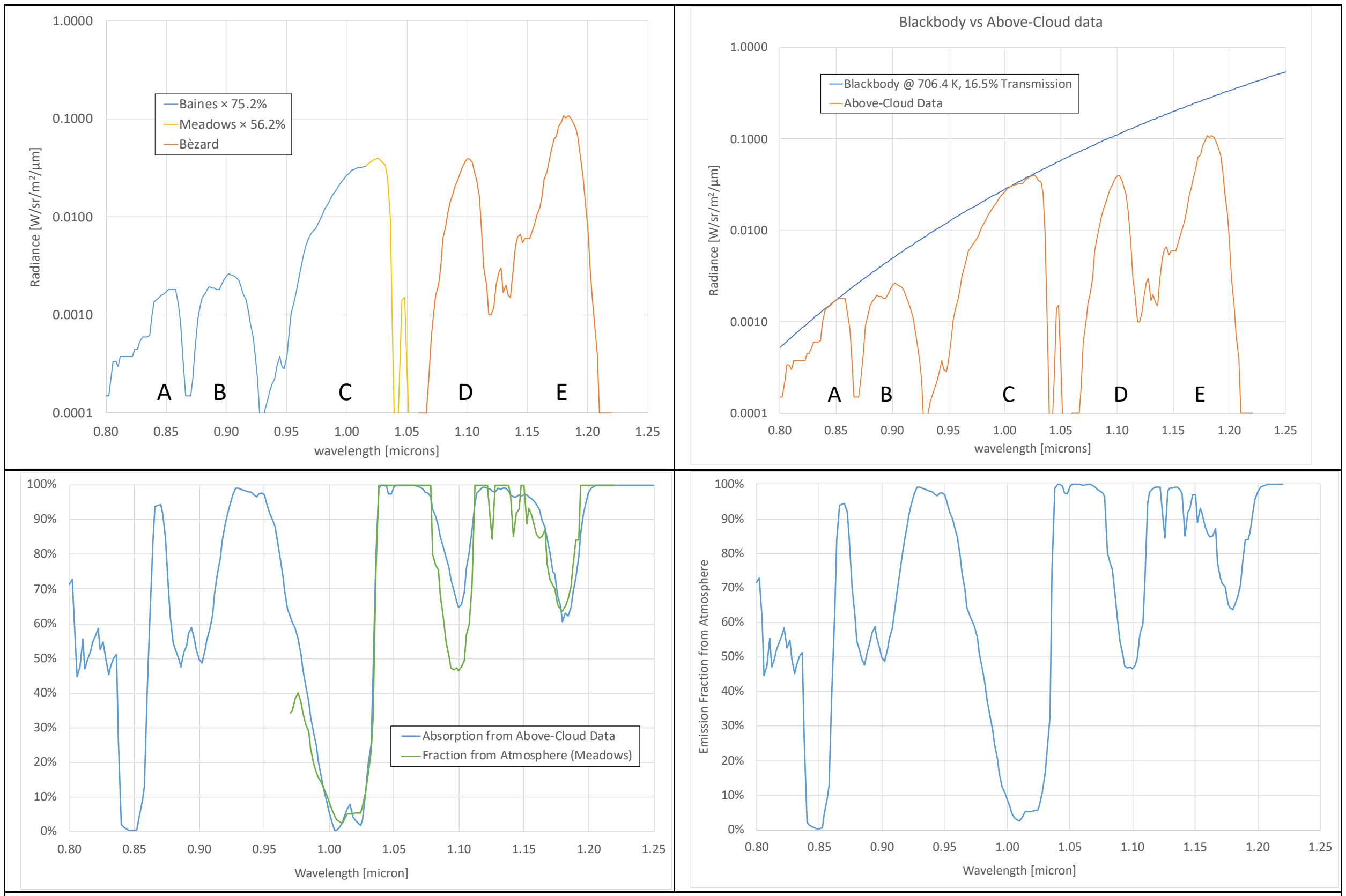

*Figure 4. (upper left) Combined above-cloud data; (upper right) 706.5 K blackbody @ 16.5% transmission and above-cloud data; (lower left) "absorption" – combination of atmospheric absorption, haze, and clouds, plotted with atmospheric emission fraction; (lower right) Atmospheric emission fraction, with crossover to absorption-based estimate at 0.996 µm. Absorption-based was also used where Meadows had no resolution.*

## 3.4 Step 4 - Clouds

Since the clouds are made up of small particles, the cloud losses are mainly from Mie scatter and should be relatively flat over our narrow passband. Just as we did above for the atmospheric emission, we fit the cores of the A and C bands with a single function representing total absorption × emissivity. We find a reduced blackbody fit with 712.4 K and 13.4% "emissivity" fits well (*Figure 5*, upper left). We now assume that the real surface temperature is 735 K, with the same bulk emissivity of 0.85 as Meadows (*Figure 5*, upper left). The ratio of the two reduced blackbody curves is the cloud loss (*Figure 5,* upper right). As expected, the clouds are relatively flat, but losses are slightly more towards the blue, likely due to particle size effects. Any wavelength dependence of the surface emissivity is not separable from cloud loss, and so combined into the cloud loss estimate.

Dividing the ground and atmosphere emission by the cloud transmission gives the spectrum of the emission below the clouds (*Figure 5*, lower left). This is what a non-imaging spectrometer would see when looking down from just below the clouds during nighttime.

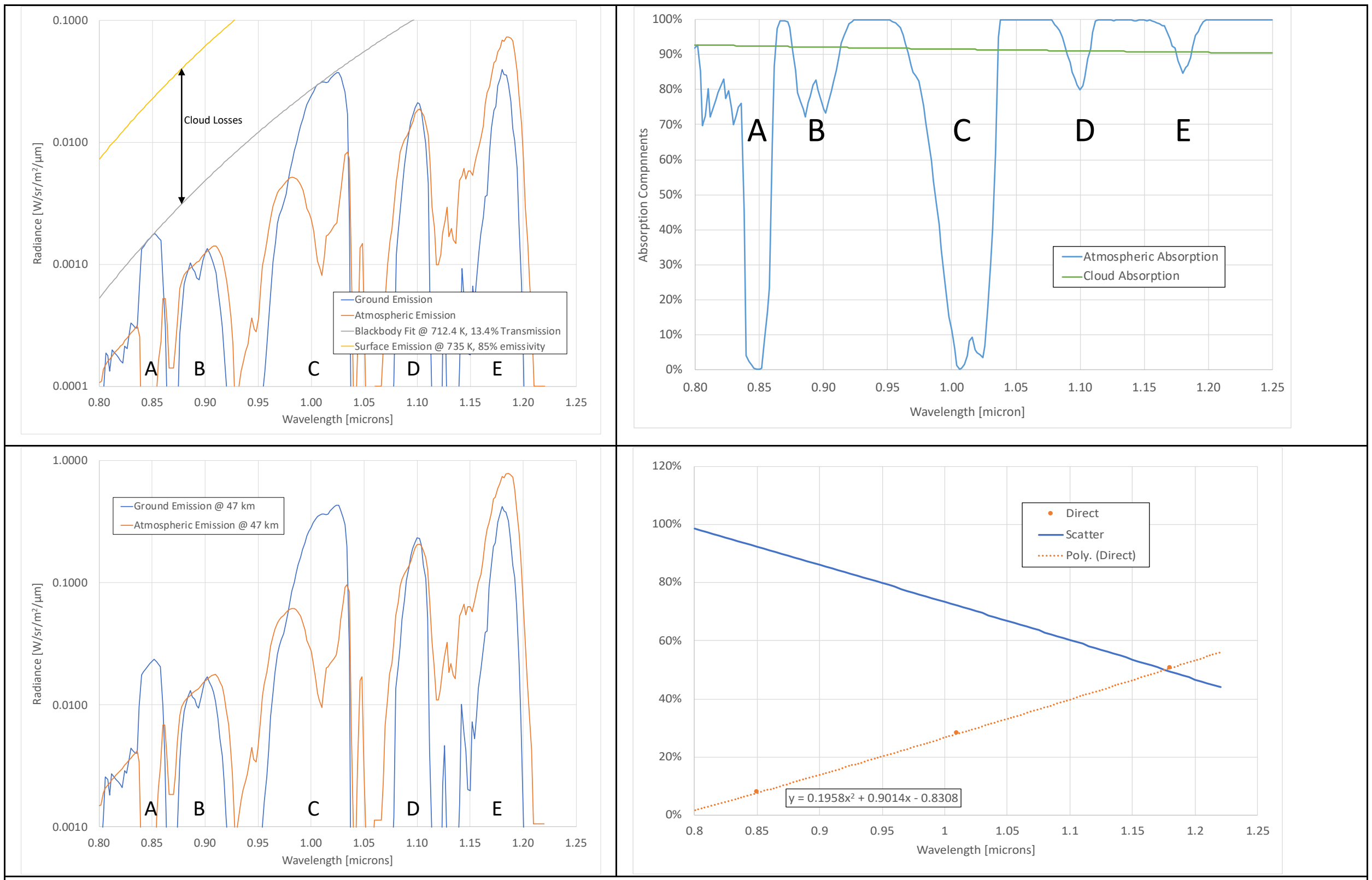

*Figure 5. (Upper left) The ratio of the fit to the observed above-cloud surface emission to the expected ground emission is the cloud loss. (Upper right) The resulting cloud absorption, plotted with the atmospheric absorption. (Lower left) ground and atmosphere emission below the clouds. (Lower right) Monte Carlo results of fraction of scattered and unscattered light from the surface.*

## 3.5 Step 5 - Rayleigh scattering from Monte Carlo simulation

Photons traveling from the ground to a camera below the cloud deck can be either absorbed or scattered. The previous sections modeled the absorption but not the scattering. Scattering below the clouds is mainly from Rayleigh scattering by $CO_2$ molecules. On Earth, Rayleigh scatter is optically thin at most visible wavelengths, but becomes significant enough in the blue to create a near-uniform "haze" that makes the entire sky appear uniformly blue. The situation on Venus is exactly the same. Unfortunately, the transition wavelength from optically thin to optically thick for Venus is right in our near-IR wavelength region; the simple analytical calculations that are available for optically thick and thin cases are not available in our passband of interest.

We have performed Monte Carlo simulations [2,6,7] specifically for the sub-cloud radiometry calculation. Davis et al. [7] contains a detailed explanation of the modeling, considers particle scatter from mid-altitude particle haze, and provides a comparison to previous scatter simulations. For the present estimations, we use the results of Davis et al. [2], which are multi-wavelength across our passband (Table 1).

The Monte Carlo simulations show that, as we expect, the scattered photons create a radiance field with very little structure, at least with the notional camera field-of-view (~10 km), so imaging resolution of the direct photons is not degraded, other than by the addition of a near-uniform background of shot noise.

**Table 1.** Results of Monte Carlo simulations from Davis et al. [2].

| **Wavelength** | $\mathbf{T_{dir}}$ | $\mathbf{F_{up}}$ | **Direct** |
|---|---|---|---|
| 0.85 μm | 6.0% | 21.9% | 7.7% |
| 1.01 μm | 24.3% | 13.0% | 27.9% |
| 1.18 μm | 46.5% | 8.0% | 50.5% |

The Monte Carlo simulation (see *Figure 6*) is run backwards, meaning that photons are traced backward starting at the camera. In the absence of atmospheric absorption, a photon received at a camera just below the cloud deck can originate from one of three kinds of history:

- Direct and uninterrupted travel from the ground. These photons can be imaged at full resolution (fraction is $T_{dir}$).
- An indirect ground-emitted photon, that is, scattered at least once by sub-cloud atmospheric Rayleigh scatter, including the possibility of being reflected back downward off the (Mie-scattering) cloud deck, or reflected off the ground (overall fraction is $T_{sca}$).
- A photon emitted by the clouds downwards and then scattered upwards (fraction is $F_{up}$). As the clouds and space above them are cold in comparison with the surface, these photons are in essence never emitted at our wavelengths.

The fractions $T_{sca}+F_{up}+T_{dir}$ add up to 1. Note that, from orbit, direct and indirect ground photons reaching cloud base look alike, since the ground resolution from orbit is 100 km. exists because the but $F_{up}$ is just divided out. We can safely assume that, inasmuch as they absorb and emit, the clouds are cold and, inasmuch as they transmit downward, outer space is much colder. We can therefore discard the $F_{up}$ term, which would otherwise add to the scatter haze and is a small correction. For photons emitted by the surface, the direct signal contribution in the rightmost column of Table 1 is then $T_{dir}/(T_{dir}+T_{sca}) = T_{dir}/(1–F_{up})$, and the complementary scattering contribution is $T_{sca}/(T_{dir}+T_{sca}) = T_{sca}/(1–F_{up})$.

The left panel of *Figure* 6 shows the atmospheric point spread function (APSF) in [km$^{-2}$]. The right panel shows the running radial integral of the APSF renormalized by $(1–F_{up})$, starting at $T_{dir}/(1–F_{up})$. Thus, it displays the fraction of the total measured radiance emitted up to a distance *r* [km] from the target point on the surface. Since the indirect/scattered light typically originates from 10s of km away from the surface point right below the sub-cloud camera, it forms a near-uniform background across the ~10-km camera field of view.

Although there is certainly some wavelength-dependent interaction between scattering and absorption, this analysis treats each separately. This interaction should have minimal effect on the final SNR, but may narrow the shapes of the atmospheric windows, since absorption is larger in the wings of the absorption bands.

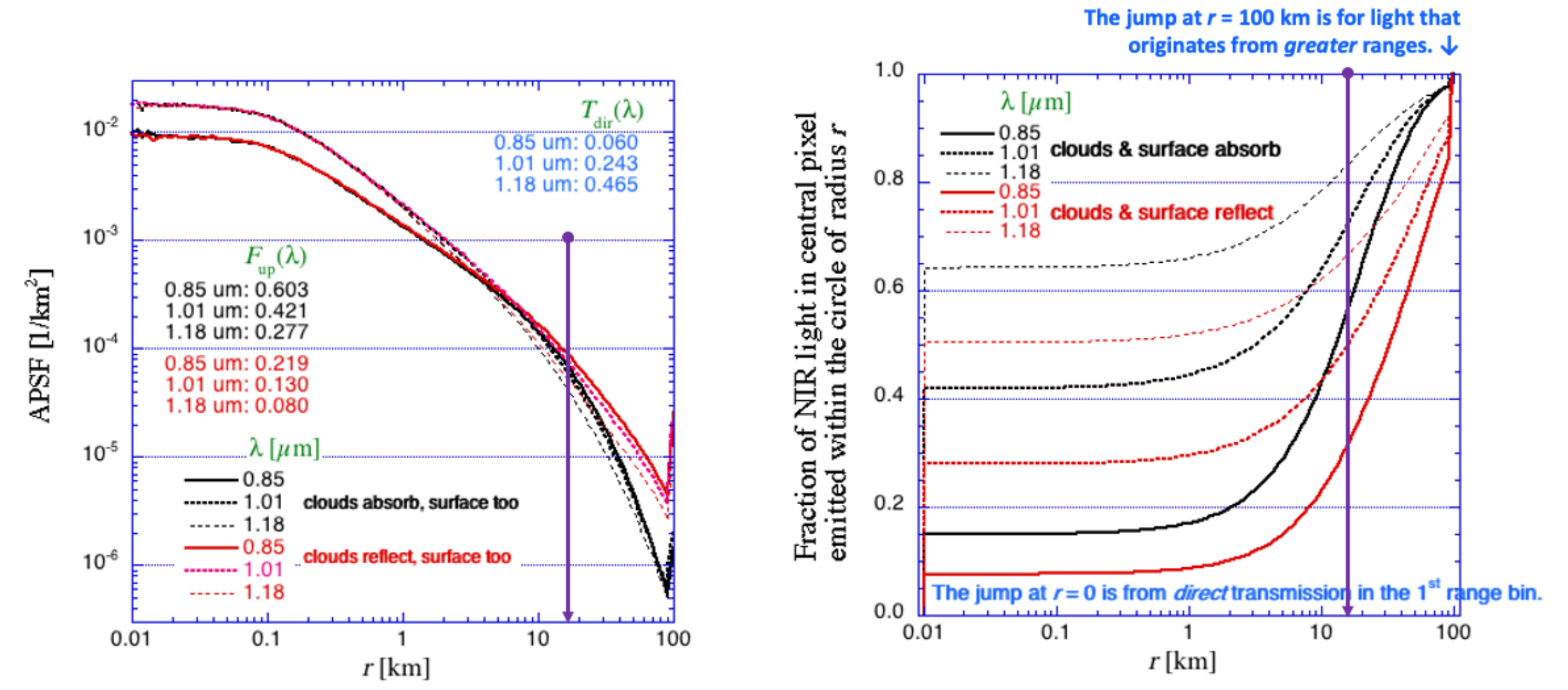

*Figure 6. Results of Monte Carlo simulations from [2]. Used here are red numbers that take albedos into account. The purple downward arrow indicates a notional half-swath for the camera (≈15 km).*

The Monte Carlo simulations were only done at three wavelengths, while this radiometry analysis is over the entire wavelength region. A quadratic polynomial fit to the simulation results are shown in (*Figure 5,* lower right).

Note that the fit becomes unphysical just below 0.8 µm. Although computing the direct component is trivial (the exponential probability of not Rayleigh scattering), computing the fraction of scattered light reaching the cloud deck is not trivial. The only realistic way to do better is to run Monte Carlo simulations for more wavelengths.

The Monte Carlos simulations give the probability that a detected photon came directly or after being scattered and/or reflected, while the previous radiometry calculations give the total. This allows separation of the scattered and direct, and thus the radiance of the direct signal and all of the noise sources. *Figure 7* (upper left) shows a summary of the radiometric components. The blue curve is the direct imaging science signal from the ground that can be used to measure emissivity and temperature. The yellow curve is the sum of atmospheric emission and scattering/reflection background that both add shot noise on top of direct ground signal. This plot should make evident that the A & B bands are challenging for a given SNR when compared to bands C, D, & E.

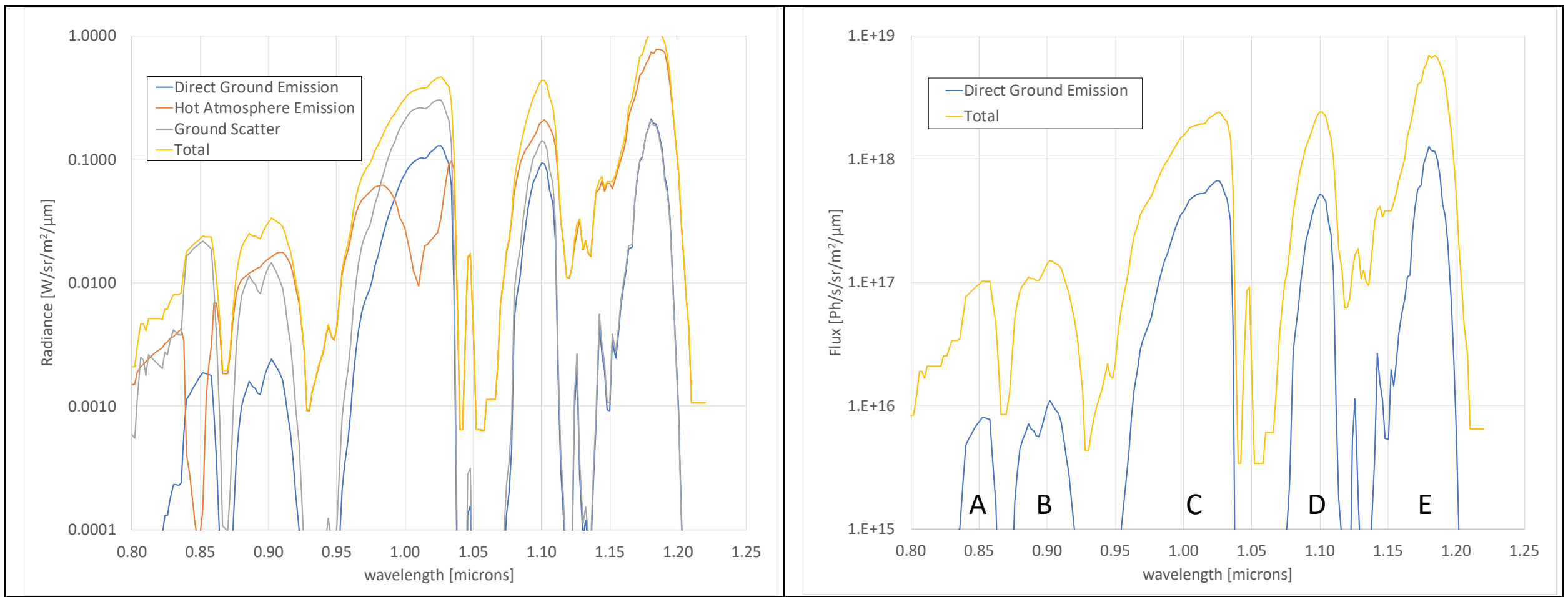

*Figure 7. (upper left) Summary of the science signal ("direct") and noise contributions. (upper right) Direct and total, converted to photon flux.*

### 3.6 Step 6 – Optimal filter passbands

Detectors in the near-IR are photon counting. *Figure 7* (upper right) shows the photon flux of the signal and background. Since the background is much brighter than the signal, the noise, and thus the SNR, is dominated by background shot noise. For each band, most of the signal comes from the core, while the wings gradually contribute more noise than signal. Table 2 shows the optimal band center and width for each band, as well as the total band photon flux. Note that, if the instrument QE is changing rapidly across the passband (e.g., Silicon detectors at 1.01 µm), then the optimal passband edges will shift accordingly.

Table 2. Optimal filter passbands to minimize shot noise

| Term | Units | A | B | C | D | E |
|---|---|---|---|---|---|---|
| Direct Sum | Ph/s/sr/m² | 1.49E+14 | 2.70E+14 | 2.29E+16 | 8.29E+15 | 2.02E+16 |
| Total Sum | Ph/s/sr/m² | 2.11E+15 | 4.64E+15 | 9.21E+16 | 4.39E+16 | 1.40E+17 |
| Bandwidth | µm | 0.024 | 0.042 | 0.056 | 0.026 | 0.028 |
| Center | µm | 0.849 | 0.896 | 1.007 | 1.098 | 1.181 |

## 4. Camera Performance

### 4.1 Step 1 – Signal-to-Noise Ratio

We would now like to know the minimum ground sample distance (GSD) that a camera can achieve for a given band and signal-to-noise ratio (SNR). GSD is distance between the samples (pixels) on the ground. The resolution (smallest detectable feature) would be some value larger than twice the GSD, due to the Nyquist limit. We discuss the resolution in the following section.

The photon fluxes are in units of Ph/s/sr/m². To get photon counts, we multiply the photon fluxes by the exposure time $T$ in seconds, the solid angle Ω of the patch of ground in steradians, and the effective aperture area of the camera $A_{\mathrm{eff}}$ in m². "Effective" here means that the true aperture area is decreased by detector quantum efficiency (QE) effects and lens throughput. As an example, a camera with a 50-mm diameter aperture lens, a 70% QE and a 70% throughput would lead to $A_{\mathrm{eff}} = \pi\,(0.05/2)^2 \times 0.7 \times 0.7 = 9.6 \times 10^{-4}$ m².

We levy the requirement that a camera can image the ground with a shot-noise-limited SNR of 100:1 per ground sample. Such a camera would be limited by systematics, rather than noise. An SNR of 100:1 corresponds to measuring the emissivity of the surface to 1% (1σ). Since detectors in the near-IR region generally use photon counting, the dominant noise source is shot noise. Defining $F_D$ as the photon flux directly from the surface and $F_T$ as the total photon flux, the SNR is then

$$R = \frac{F_D T \Omega A_{\mathrm{eff}}}{(F_T T \Omega A_{\mathrm{eff}})^{1/2}}.$$

We want to write an equation to solve for the minimum GSD that achieves our target SNR. We denote this minimum GSD as $G$, in units of meters. The camera target distance $D$ corresponds to an altitude of $D$ = 47,000 m. The solid angle of the square patch on the ground as seen by the camera is then $\Omega = (G/D)^2$ in steradians.

A probe at 47-km altitude would be traveling horizontally with the wind, assumed hereafter to nominally be $V$ = 60 m/s. We limit smear of the image due to wind to be some fraction $S$ of the GSD. Setting $S$ equal to 1 is an equipartition of contributions to the resolution budget between motion smear and square "pixels", discussed in the following section. With this smear limit, the exposure time must be less than $T = SG/V$.

Substituting in for Ω and $T$, and solving for $G$, yields

$$G = \left(\frac{F_T V}{S A_{\mathrm{eff}}}\right)^{1/3} \left(\frac{R D}{F_D}\right)^{2/3}.$$

The results for minimum ground sampling are tabulated in Table 3 for various choices of effective aperture areas $A_{\mathrm{eff}}$.

Table 3. Ground sampling achieved for SNR 100:1 per sample for various effective aperture sizes $A_{\mathrm{eff}}$. Effective aperture range corresponds to 1.6 to 9 [cm] diameter for QE × throughput of 50%.

| $A_{\mathrm{eff}}$ [m²] | A | B | C | D | E |
|---|---|---|---|---|---|
| 1.0E-04 | 108 | 94 | 13 | 20 | 17 |
| 2.0E-04 | 86 | 75 | 11 | 16 | 13 |
| 4.0E-04 | 68 | 59 | 8 | 13 | 10 |
| 8.0E-04 | 54 | 47 | 7 | 10 | 8 |
| 1.6E-03 | 43 | 37 | 5 | 8 | 7 |
| 3.2E-03 | 34 | 30 | 4 | 6 | 5 |

### 4.2 Step 2 – Resolution

SNR per ground sample is not the only camera performance factor in determining optimal sampling. A camera system with poor performance will not allow reconstruction of the target ground features. The approach we will use here to estimate resolution is to examine loss of contrast, or modulation transfer. The Modulation Transfer Function (MTF) is the Fourier transform of the Point Spread Function (PSF), modulo a normalization constant. For each spatial frequency on the surface, the MTF is the loss of signal at that frequency due to lens aberrations, pixelization, diffraction, and (for this analysis) motion smear.

We choose a single effective aperture to analyze, $A_{\mathrm{eff}} = 4 \times 10^{-4}$ m² from Table 3. For a 12-µm pixel pitch and QE × throughput = 50%, this corresponds to an f/1.77 lens with an aperture diameter of 32 mm. Commercial SWIR lenses have aberrations with an RMS diameter of roughly a pixel, which will be our assumption. A plot of the resulting MTF is shown in *Figure 8* (upper left), with the horizontal units in the conventional but unintuitive cycles per pixel. *Figure* 8 (upper right) shows the same MTF, plotted in terms of ground resolution.

The Nyquist limit on these two plots roughly corresponds to two immediately adjacent pixels. An MTF of 0.1 means that if every even pixel in the image had an emissivity of zero and every odd pixel an emissivity of one, the observed camera counts would be ±5% on each side of the mean noise. The mean noise is 1% 1-σ (±0.5%), so reconstruction of the original contrast can be performed with useful certainty. We compute the quality of reconstruction as a function of spatial frequency below.

Consider a scenario where our camera takes an image every 10 meters of travel, equivalently, every 0.16 seconds. The SNR per unit spatial frequency for the 5 bands is shown in *Figure 8* (lower panel). In bands C, D, and E, the SNR is high enough to extract spatial information down to the Nyquist resolution limit of 20 meters. The way to interpret this is that, even at the Nyquist limit of 20-m resolution, the SNR for bands C, D, and E is high enough (SNR>10:1), such that the image of the ground down at Nyquist spatial resolution can be reconstructed from the data.

For bands A and B, 20-m resolution is a lost cause, but 80-m resolution at an SNR > 100:1 is achieved by binning pixels 4x4 and co-adding 3 or more exposures. The APSFs in *Figure 6* add a near-uniform background on top of the camera MTF but, as this near-uniform background has no spatial information until we get to very large angles/distances (comparable to or exceeding the camera field-of-view), the only contribution is background shot noise. This background defines the noise floor of *Figure 8* (lower panel).

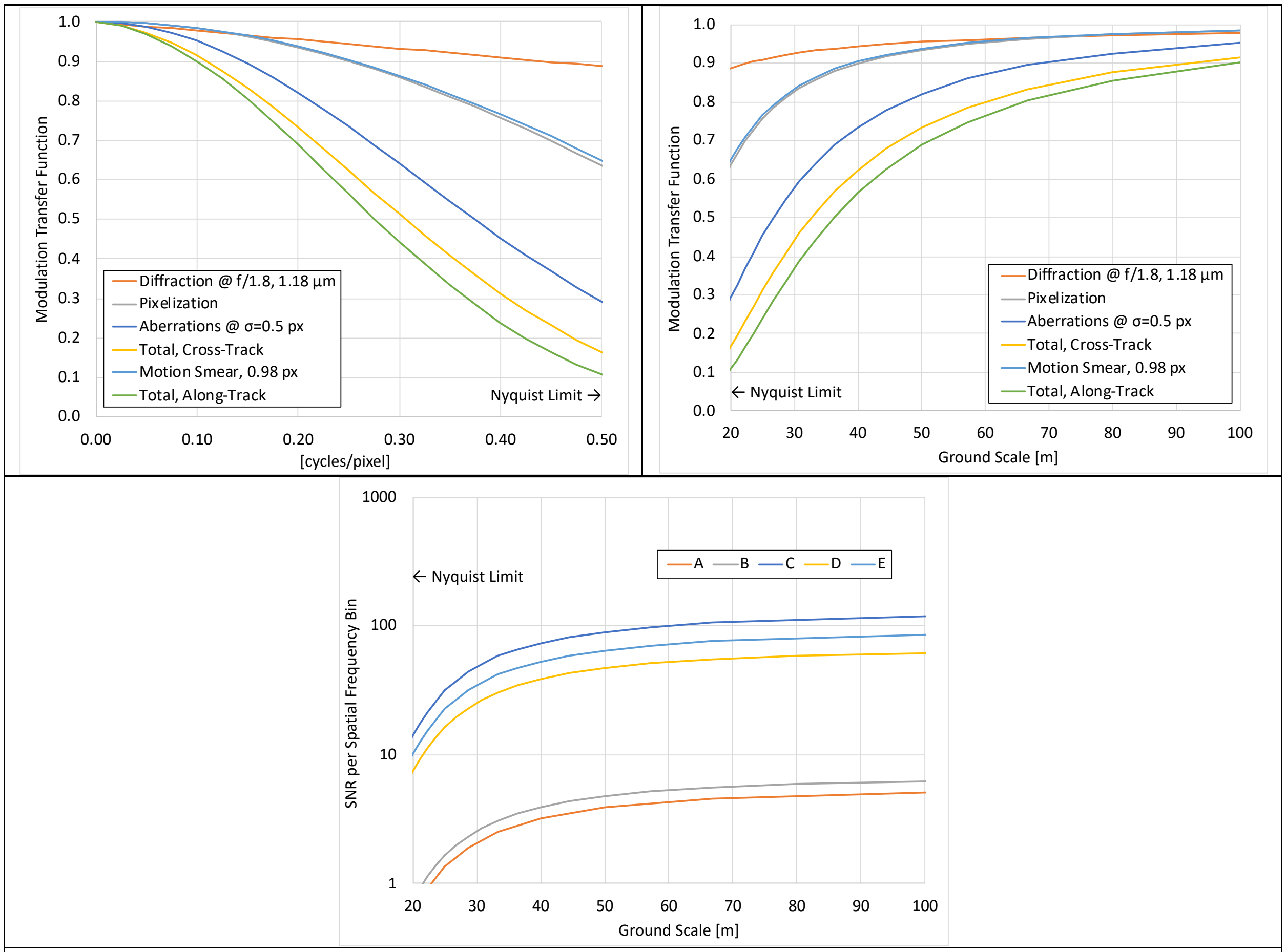

Figure 8. (Upper left) MTF of a plausible sub-cloud imaging camera. Along-track MTF has motion smear of 0.98 pixels rather than 1 so that the curve does not overlap pixelization. (Upper right) MTF, plotted in terms of resolution at the ground. (Lower panel) Single-frame along-track SNR for a camera taking exposures every 10 meters. Cross-track is not shown. Bands A and B need binning and co-adding to reach useful SNR.

## 5. Conclusions

In 2002, Moroz [10] was the first author to investigate the imaging of Venus' surface from anywhere below its clouds at visible and near-IR wavelengths, day or night. Like us, he based his estimations on the relative amounts of direct and background light. Moroz concludes that in daytime and at night for the shorter (visible) wavelengths where Rayleigh scattering is intense, the background is overwhelming at cloud base, so the camera has to be lowered into the hot near-surface atmosphere, which is not a practical proposition. Also, when there is an optically thick sub-cloud haze, direct light to cloud base is depleted too much, so again the camera needs to be lowered to image the surface. For imaging from just below the clouds, the direct/background contrast ratio we obtain—from a far more detailed study—is in rough quantitative agreement with the corresponding estimate by Moroz. We therefore concur with him about the feasibility of sharp (pixel-scale) imaging of the surface emission features at 1 µm or longer wavelengths.

In 2015, Ekonomov [11] performed a numerical investigation of Venus' sub-cloud atmospheric MTF (AMTF). Paraphrasing his conclusions in our language, he determined that the AMTF has a very broad bandpass (due to the significant direct transmission), but at a reduced amplitude (due to the non-negligible near-uniform background light). Davis et al. [7] performed a far more detailed and comprehensive study of the AMTF, starting with its cosine Fourier transform, the APSF. They came to the same conclusion with a specification across near-IR spectral channels. Here, we examined how various camera MTFs will affect the imaging on top of the AMTF. They are found to be important at pixel scales, but correctable, hence not show-stoppers.

## Acknowledgements

This research was carried out at the Jet Propulsion Laboratory, California Institute of Technology, under a contract with the National Aeronautics and Space Administration (80NM0018D0004). We thank Paul Byrne for fruitful conversations about the potential scientific benefits of this technology.